\documentclass[conference]{sig-alternate-05-2015}

\usepackage[T1]{fontenc}

\usepackage{amsmath}
\usepackage{amssymb}
\usepackage{amsfonts}
\usepackage{epsfig}
\usepackage{graphicx}
\usepackage{subfigure}
\usepackage{url}
\usepackage[bottom]{footmisc}
\usepackage{balance}
\usepackage{algorithmic}
\usepackage{algorithm}
\usepackage{multirow}
\usepackage{xspace}
\usepackage{times}
\usepackage{cite}
\usepackage[singlelinecheck=off]{caption}
\DeclareCaptionType{copyrightbox}

\newcommand{\squishlist}{
 \begin{list}{$\bullet$}
  {  \setlength{\itemsep}{0pt}
     \setlength{\parsep}{3pt}
     \setlength{\topsep}{3pt}
     \setlength{\partopsep}{0pt}
     \setlength{\leftmargin}{2em}
     \setlength{\labelwidth}{1.5em}
     \setlength{\labelsep}{0.5em}
} }
\newcommand{\squishlisttight}{
 \begin{list}{$\bullet$}
  { \setlength{\itemsep}{0pt}
    \setlength{\parsep}{0pt}
    \setlength{\topsep}{0pt}
    \setlength{\partopsep}{0pt}
    \setlength{\leftmargin}{2em}
    \setlength{\labelwidth}{1.5em}
    \setlength{\labelsep}{0.5em}
} }

\newcommand{\squishdesc}{
 \begin{list}{}
  {  \setlength{\itemsep}{0pt}
     \setlength{\parsep}{3pt}
     \setlength{\topsep}{3pt}
     \setlength{\partopsep}{0pt}
     \setlength{\leftmargin}{1em}
     \setlength{\labelwidth}{1.5em}
     \setlength{\labelsep}{0.5em}
} }

\newcommand{\squishend}{
  \end{list}
}

\newcommand{\eat}[1]{}

\newcommand{\NP}{\ensuremath{\mathbf{NP}}\xspace}

\newcounter{ccc}

\newcommand{\bigO}{\mathcal{O}}

\begin{document}
\setcopyright{acmcopyright}




\acmPrice{\$15.00}

%

\title{Vertex-Centric Graph Processing: \\ The Good, the Bad, and the Ugly}

%
%
%
%
%

\numberofauthors{1} 
%
\author{\alignauthor Arijit Khan\\
\affaddr{Nanyang Technological University, Singapore}\\
\email{arijit.khan@ntu.edu.sg}
}

\maketitle
\pagenumbering{arabic}

\vspace{-15mm}
\begin{abstract}
We study distributed graph algorithms that adopt an iterative vertex-centric framework for graph processing,
popularized by the Google's Pregel system. Since then, there are several attempts to implement many graph algorithms
in a vertex-centric framework, as well as efforts to design optimization techniques for improving the efficiency.
However, to the best of our knowledge, there has not been any systematic study to compare these vertex-centric implementations
with their sequential counterparts.  Our paper addresses this gap in
two ways. (1) We analyze the
computational complexity of such implementations with the notion of time-processor product,
and benchmark several vertex-centric graph algorithms whether they perform more work with
respect to their best known sequential solutions. (2) Employing the concept of balanced practical Pregel algorithms,
we study if these implementations suffer from imbalanced workload and large number of iterations.
Our findings illustrate that with the exception of Euler tour tree algorithm, all other algorithms either perform
more work than their best-known sequential approach, or suffer from imbalanced workload/ large number of iterations,
or even both. We also emphasize on graph algorithms that are fundamentally difficult to be expressed
in vertex-centric frameworks, and conclude by discussing the road ahead for distributed graph
processing.
\end{abstract}

\section{Introduction}
\label{sec:intro}

In order to achieve low latency and high throughput
over massive graph datasets, distributed solutions were proposed in which the
graph and its data are partitioned horizontally across cheap commodity
servers in the cluster. The distributed programming model for large graphs
has been popularized by the Google's Pregel framework \cite{MABDHLC10}.
It hides distribution related details such as data partitioning,
communication, underlying system architecture,
and fault tolerance behind an abstract API. Also
known as the {\em think-like-a-vertex} model, it requires that the user
expresses the computation from the perspective of a single vertex,
by providing a higher-order vertex-compute() function.

\vspace{0.1mm}

In Pregel, which was inspired by the Bulk Synchronous Parallel (BSP)
model \cite{V90}, graph algorithms are expressed as a sequence
of iterations called supersteps. 
Each superstep is an atomic
unit of parallel computation. During a superstep, Pregel executes
a user-defined function for each vertex in parallel. The user-defined
function specifies the operation at a single vertex $v$ and at a single
superstep $S$. The supersteps are globally synchronous among all
vertices, and messages are usually sent along the outgoing edges
from each vertex. In 2012, Yahoo! launched the Apache Giraph
as an open-source project, which clones the concepts of Pregel.
%

\vspace{0.1mm}

With the inception of the Pregel framework, vertex-centric distributed graph processing
has become a hot topic in the database community (for a
survey, see \cite{KE14,YBTDC16,MWM15,HDAOWJ14}). Although Pregel
provides a high-level distributed programming abstract, it suffers from efficiency issues such as
the overhead of global synchronization, large volume of messages, imbalanced
workload, and straggler problem due to slower machines. Therefore, more advanced
vertex-centric models (and its variants) have been proposed, e.g., asynchronous (GraphLab),
asynchronous parallel (GRACE), barrierless asynchronous parallel (Giraph Unchained),
gather-apply-scatter (PowerGraph), timely dataflow (Naiad),
data parallel (GraphX, Pregelix),
and subgraph centric frameworks (NScale, Giraph++).
Various algorithmic and system-specific optimization techniques were also designed,
e.g., graph partitioning and re-partitioning, combiners and aggregators, vertex scheduling, superstep sharing, message reduction,
finishing computations serially,
among many others.

\vspace{0.1mm}

While speeding up any algorithm is always significant in its own right, there may be circumstances in which we would not benefit
greatly from doing so. McSherry et. al. \cite{MIM15} empirically demonstrated that single-threaded implementations
of many graph algorithms using a high-end 2014 laptop are often an order of magnitude faster than the published results for
state-of-the-art distributed graph processing systems using multiple commodity machines and hundreds of cores over the same
datasets. Surprisingly, with the exception of \cite{YCXLNB14}, the complexity of vertex-centric graph algorithms
has never been formally analyzed. As one may realize, this is not a trivial problem --- there are multiple
factors involved in a distributed environment including the number of processors, computation time,
network bandwidth, communication volume, and memory usage. To this end, we make the following contributions.
\begin{itemize}
\item We formally analyze the computational complexity of vertex-centric implementations with the notion of time-processor product \cite{V90},
and benchmark several vertex-centric graph algorithms whether they perform more work
in comparison to their best known sequential algorithms.
\item We employ the concept of balanced, practical Pregel algorithms (BPPA) \cite{YCXLNB14} to investigate if these vertex-centric graph algorithms
suffer from imbalanced workload and large number of iterations.
\end{itemize}

While the notion of balanced, practical Pregel algorithms was introduced by Yan et. al. \cite{YCXLNB14}, they only considered
the connected component-based algorithms. On the contrary, in this paper
we study as many as twenty different graph algorithms (Table~\ref{tab:efficiency}), whose vertex-centric algorithms were implemented in the literature.
Finally, we also identify graph workloads and algorithms that are difficult to be expressed in the vertex-centric framework,
and highlight some important research directions.

\section{Preliminaries}
\label{sec:preliminaries}

In the following, we introduce two metrics: time-processor product and balanced, practical Pregel algorithms. The first
one is used to measure if a vertex-centric algorithm performs more work compared to the problem's best known sequential solution.
We consider the second metric to verify if a vertex-centric implementation suffers from imbalanced workload and large number of iterations.

\footnotetext[1]{\scriptsize For higher values of $g$, the time-processor product would be even higher.}

\subsection{Time-Processor Product}
\label{sec:time_processor}

Time-processor product was employed by Valiant \cite{V90}
as a complexity measure of algorithms on the BSP model, which is defined by the following parameters. (1) Bandwidth parameter
$g$, that measures the permeability of the network to continuously send traffics to uniformly-random destinations.
The parameter $g$ is defined such that an $h$-relation will be delivered in time $hg$.
The value of $g$ is normalized with respect to the clock rate of each architecture
so that it is in the same units as the time for executing sequences of instructions. (2) Synchronization periodicity $L$,
where the components at regular intervals of $L$ time units are synchronized. In a superstep of periodicity $L$, $L$ local
operations and $\lfloor L/g \rfloor$-relation message patterns can be realized. (3) The number of processors $p$.
The time charged for a superstep is calculated as follows. Let $w_i$ be the amount of local work
performed by processor $i$ in a given superstep. Assume $s_i$ and $r_i$ be the number of messages sent and received, respectively,
by processor $i$. Let $w = \max_{i=1}^p w_i$, and $h=\max_{i=1}^p(\max(s_i,r_i))$. Then, the time for a superstep is given by $\max(w,gh,L)$.

\vspace{0.1mm}

If we have multiple processors at our disposal, we can solve a problem more quickly by dividing it into independent sub-problems and
solving them at the same time, one at each processor. The running time of the algorithm is then the longest running time of any
of these processors. More specifically, given an input size $n$, the running time $T(n)$ is the elapsed time from
when the first processor begins executing to when the last processor stops executing. A BSP algorithm for a given problem
is called efficient if its processor bound $P(n)$ and time bound $T(n)$ are such that time-processor product $P(n)T(n)=\bigO(S)$,
where $S$ is the running time of the best known sequential algorithm for the problem, provided that $L$ and $g$ are below
certain critical values. Therefore, with this metric, we measure whether a vertex-centric algorithm performs more work, compared
to the problem's best-known sequential algorithm.

\subsection{Balanced, Practical Pregel Algorithms}
\label{sec:bppa}

For an undirected graph, let us denote by $d(v)$ the degree of vertex $v$. On the other hand,
let $d_{in}(v)$ and $d_{out}(v)$ denote the in-degree and out-degree, respectively, of vertex $v$
in a directed graph. A Pregel algorithm is called a balanced,
practical Pregel algorithm (BPPA) \cite{YCXLNB14} if it satisfies the following properties.
(1) Each vertex $v$ uses $\bigO(d(v))$ (or, $\bigO(d_{in}(v)+d_{out}(v))$) space of storage.
(2) The time complexity of the vertex-compute() function for each vertex $v$ is $\bigO(d(v))$ (or, $\bigO(d_{in}(v)+d_{out}(v))$).
(3) At each superstep, the size of the messages sent/received by each vertex $v$ is $\bigO(d(v))$ (or, $\bigO(d_{in}(v) + d_{out}(v))$).
(4) The algorithm terminates after $\bigO(\log n)$ supersteps.
Properties 1-3 offers good load balancing and linear cost at each
superstep, whereas property 4 impacts the total running time.
\begin{figure}[tb!]
\centering
\subfigure [\scriptsize Superstep 0] {
\includegraphics[scale=0.16]{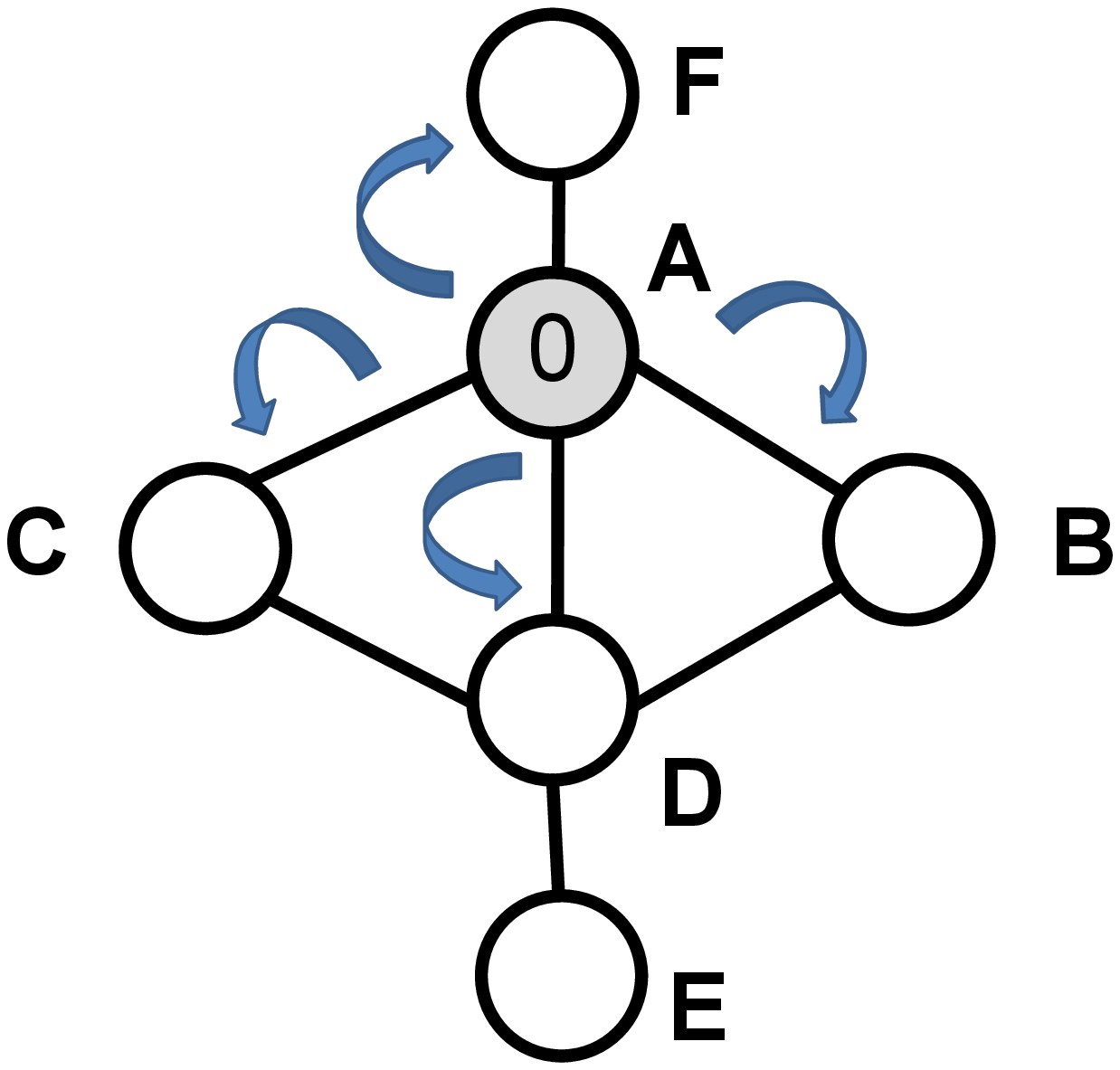}
\label{fig:diameter1}
}
\subfigure [\scriptsize Superstep 1] {
\includegraphics[scale=0.16]{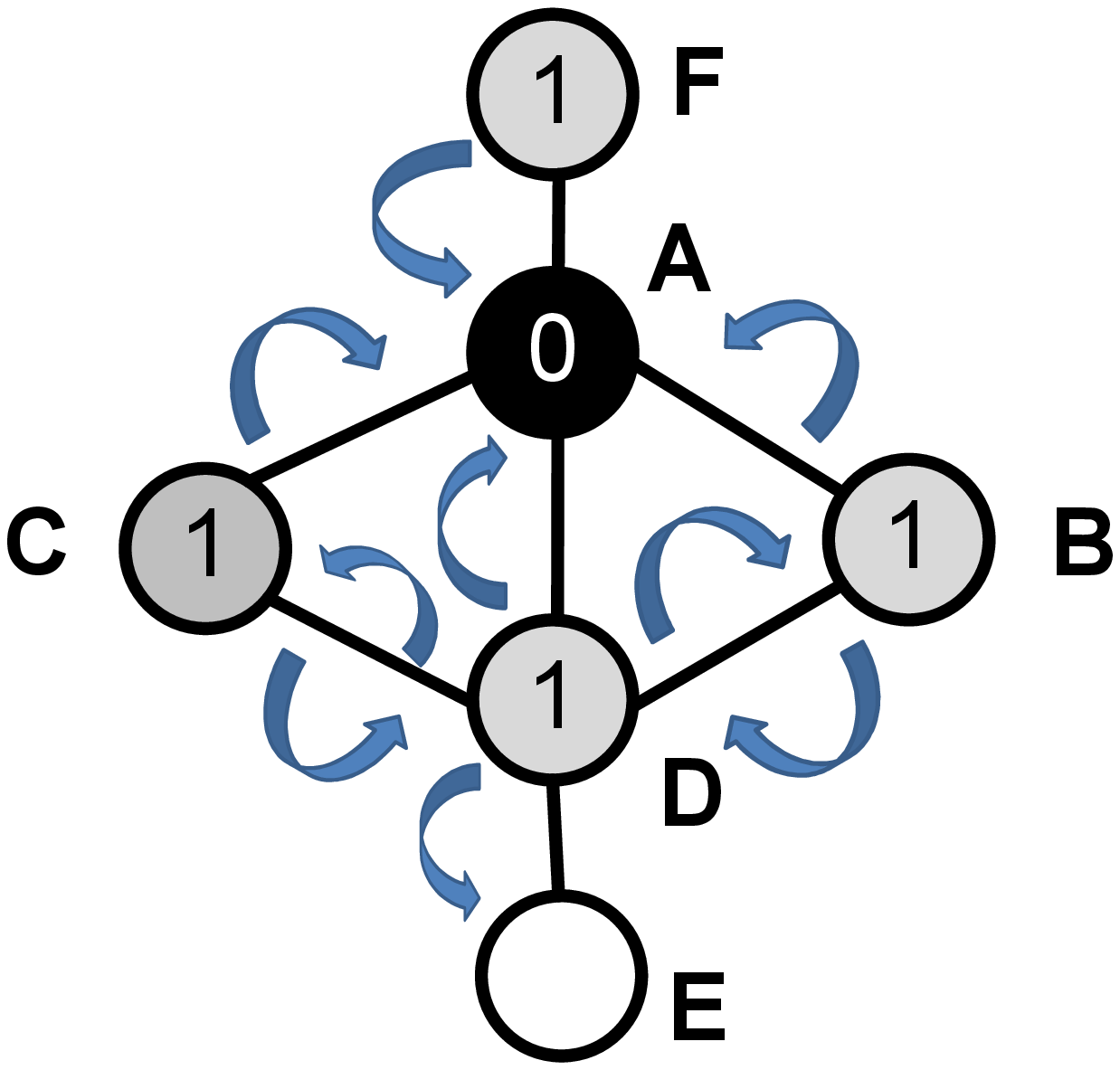}
\label{fig:diameter2}
}
\subfigure [\scriptsize Superstep 2] {
\includegraphics[scale=0.16]{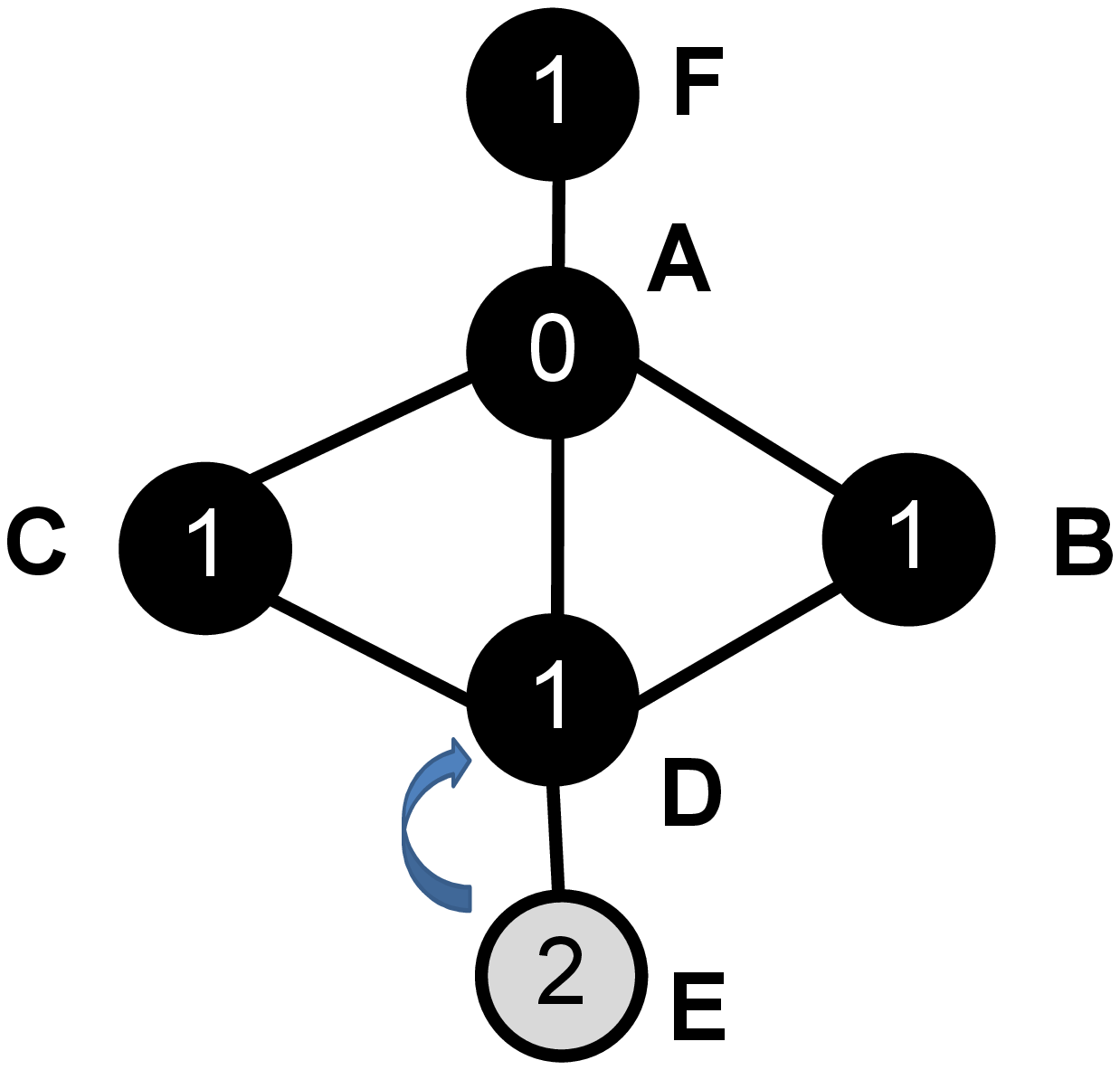}
\label{fig:diameter3}
}
\vspace{-3mm}
\caption{\scriptsize Vertex-centric algorithm for diameter computation in unweighted graphs}
\label{fig:diameter}
\vspace{-4mm}
\end{figure}

\begin{table*} [ht]
\scriptsize
\centering
\vspace{-2mm}
\begin{tabular} { |l|c||c|c||c|c||c|c| }
\hline
& \textsf{Graph} & \multicolumn{2}{c||}{\textsf{Vertex-Centric}} & \multicolumn{2}{c||}{\textsf{Best Sequential}} & \multicolumn{2}{c|}{\textsf{Vertex-Centric}}  \\
& \textsf{Workload} & \textsf{Algorithm} & \textsf{Complexity} & \textsf{Algorithm} & \textsf{Complexity} & \textsf{More Work?} & \textsf{BPPA?} \\ \hline \hline
1& Diameter (Unweighted) & \cite{PW15} & $\bigO(mn)$ & BFS \cite{RW13}                    & $\bigO(mn)$  & {\bf No}  & No  \\ \hline
2& PageRank \footnotemark[2]              & \cite{MABDHLC10} & $\bigO(mK)$ & power iteration & $\bigO(mK)$  & {\bf No}  & No  \\ \hline
3& Connected Component   & Hash-Min \cite{MABDHLC10} & $\bigO(m\delta)$ & BFS \cite{HT73}           & $\bigO(m+n)$ & Yes & No  \\ \hline
4& Connected Component   & S-V \cite{YCXLNB14} & $\bigO((m+n)\log n)$ & BFS \cite{HT73} & $\bigO(m+n)$ & Yes & No \\ \hline
5& Bi-Connected Component& \cite{YCXLNB14} & $\bigO((m+n)\log n)$ & DFS \cite{HT73}                & $\bigO(m+n)$ & Yes & No \\ \hline
6& Weakly Connected Component& \cite{YCXLNB14} & $\bigO((m+n)\log n)$ & BFS \cite{HT73}            & $\bigO(m+n)$ & Yes & No \\ \hline
7& Strongly Connected Component & \cite{YCXLNB14} & $\bigO((m+n)\log n)$ & DFS \cite{T72}          & $\bigO(m+n)$ & Yes & No \\ \hline
8& {\bf Euler Tour of Tree} & \cite{YCXLNB14} & $\bigO(n)$ &       DFS                         & $\bigO(n)$   & {\bf No}& {\bf Yes} \\ \hline
9& Pre- \& Post-order Tree Traversal& \cite{YCXLNB14} & $\bigO(n\log n)$  &   DFS              & $\bigO(n)$   & Yes & {\bf Yes} \\ \hline
10& Spanning Tree        & \cite{YCXLNB14, TV84}  & $\bigO((m+n)\log n)$  & BFS                                & $\bigO(m+n)$ & Yes & No \\ \hline
11& Minimum Cost Spanning Tree \footnotemark[2] & \cite{SW14} & $\bigO(\delta m\log n)$ & Chazelle's algorithm \cite{C00}    & $\bigO(m\alpha(m,n))$ & Yes & No \\ \hline
12& Graph Coloring with   &  \cite{SW14}  & $\bigO(Km\log n)$ &  Lexicographically First  &  $\bigO(Km)$  &  Yes   &  No    \\
  & Maximal Independent Set \footnotemark[2] &  &     &    Maximal Independent Set               &                &        &      \\ \hline
13& Maximum Weight Matching & \cite{SW14} & $\bigO(Km)$ & Pries Algorithm \cite{P99} & $\bigO(m)$ & Yes & No     \\
  & with Pries Algorithm \footnotemark[2] &     &       &                                    &              &     &      \\ \hline
14& Bipartite Maximal Matching &  \cite{MABDHLC10} & $\bigO(m\log n)$ & greedy & $\bigO(m+n)$ & Yes & {\bf Yes} \\
  & (Unweighted)               &     &     &  & &  & \\ \hline
15& Betweenness Centrality & \cite{RSP13} & $\bigO(mn)$ & Brandes' algorithm \cite{B01} & $\bigO(mn)$ & {\bf No} & No  \\
  & (Unweighted)           & & &                               &             &  &     \\ \hline
16& Single-Source Shortest Path & \cite{MABDHLC10} & $\bigO(mn)$ & Dijkstra with Fibonacci heap & $\bigO(m + n\log n)$ & Yes & No \\ \hline
17& All-pair Shortest Paths & \cite{PW15} & $\bigO(mn)$ & Chan's algoithm \cite{C12}  & $\bigO(mn)$  & {\bf No}  & No   \\
& (Unweighted)              & & &                             &              &     &      \\ \hline
18& Graph Simulation \footnotemark[2] & \cite{FNRMS13} & $\bigO(m^2(n_q+m_q))$ & Henzinger et. al. \cite{HHK95} & $\bigO\left(\left(m+n\right)\left(m_q+n_q\right)\right)$ & Yes &   No   \\ \hline
19& Dual Simulation \footnotemark[2]  & \cite{FNRMS13} & $\bigO(m^2(n_q+m_q))$ & Ma et. al. \cite{MCFHW14} & $\bigO\left(\left(m+n\right)\left(m_q+n_q\right)\right)$ &  Yes   &   No   \\ \hline
20& Strong Simulation \footnotemark[2]& \cite{FNRMS13} & $\bigO(m^2n(n_q+m_q))$ & Ma et. al. \cite{MCFHW14} & $\bigO\left(n\left(m+n\right)\left(m_q+n_q\right)\right)$ & Yes & No \\ \hline
\hline
\end{tabular}
\vspace{-2mm}
\caption{\small Efficiency benchmark for vertex-centric graph algorithms: \# nodes = $n$, \# edges = $m$, diameter = $\delta$ \label{tab:efficiency}}
\vspace{-2mm}
\end{table*}

\section{Complexity Benchmark}
\label{sec:benchmark}

We summarize our complexity benchmark for twenty vertex-centric graph algorithms in Table~\ref{tab:efficiency}.

\subsection{Diameter Computation}

We consider a vertex-centric algorithm \cite{PW15} that computes the exact diameter of an unweighted graph.
Let us denote the eccentricity $\epsilon(v)$ of a vertex $v$ as the largest hop-count distance from $v$ to any other
vertex in the graph. The diameter $\delta$ of the graph is defined as the maximum eccentricity over all its nodes.
Instead of finding this largest vertex eccentricity one-by-one, the algorithm works by computing the eccentricity of
all vertices simultaneously.

\vspace{0.1mm}

We illustrate in Figure~\ref{fig:diameter} the eccentricity computation method of one
vertex. Initially, each vertex adds it's own unique id to the outgoing messages (sent along the outgoing edges)
and also to the history set, which resides in the local memory of that vertex. After the initial superstep, the algorithm operates by iterating through
the set of received ids, which correspond to the vertex that sent the original message. The receiving vertex then constructs
a set of outgoing messages by adding each element of the incoming set which was not seen yet. The reason
for keeping a history of the originating ids that were received earlier is to prevent the re-propagation of a message
to the same vertices. The history set also serves to prune the set of total messages
by eliminating message paths that would never result in the vertex's eccentricity.

\vspace{0.1mm}

All vertices originate a unique message in superstep 1, and maintain a history of which messages
they have and have not received, and the algorithm continues until there are no more messages to propagate.
Assuming the graph is connected, each vertex will process a message from each originating vertex
exactly once. The algorithm terminates when the largest eccentricity is calculated; and therefore, the diameter of the graph
is equal to the number of supersteps (minus 1, for the final, non-processing superstep).

\vspace{0.1mm}

Since each vertex generates a unique message, there are  $\Theta(n)$ messages present in the
graph. Each message will be passed $\bigO(m)$ times, resulting in a total message complexity of
$\bigO(mn)$. There will be total $\bigO(\delta)$ supersteps. Each vertex also processes $n$ messages;
therefore, the overall computation cost is $\bigO(n^2)$. Assuming bandwidth parameter \footnotemark[1] $g=\bigO(1)$,
the time-processor product = $\bigO(mn)$, which is equal
to the complexity of the best-known sequential algorithm.

\vspace{0.1mm}

However, this vertex-centric algorithm is not BPPA due to the following reasons.
(1) The number of messages that each vertex $v$ relays can be asymptotically
larger than $\bigO(d(v))$ at later supersteps.
(2) Given that each vertex $v$ must store a history of the messages
received, each vertex stores $\bigO(n)$ vertex IDs, which
is larger than $\bigO(d(v))$.
(3) There are total $\bigO(\delta)$ supersteps, which could be larger than $\bigO(\log n)$.

\vspace{0.1mm}

One may note that the above vertex-centric algorithm also computes all-pair-shortest-paths (APSP) in an unweighted
graph. Therefore,
APSP computation using the above implementation has the same complexity, as
presented in Table~\ref{tab:efficiency}.

\subsection{PageRank}

A vertex-centric implementation of the PageRank algorithm is given in the original Pregel paper \cite{MABDHLC10}.
At superstep 0, the PageRank value of each vertex is $\frac{1}{n}$. In every remaining superstep until
convergence, each vertex sends along each outgoing edge its tentative PageRank divided by the number of outgoing edges.
Starting from superstep 1, each vertex sums up the values arriving on messages into sum and sets its own PageRank
to $\frac{(1-\alpha)}{n} + \alpha \times sum$, where $\alpha$ is a constant teleportation probability.
After reaching convergence (or, a predefined number of supersteps), no further messages are sent
and each vertex votes to halt.

\vspace{0.1mm}

At each superstep, there are $\bigO(m)$ messages sent/received, and $\bigO(m)$ computations performed. For
$g=\bigO(1)$, the time-processor product = $\bigO(Km)$, where $K$ is the number of supersteps. This
matches with the complexity of the best-known sequential algorithm. Though it is a balanced Pregel algorithm
(i.e., satisfies properties 1-3), but not BPPA, since $K$ can be larger than $\bigO(\log n)$,
usually in the order of 30 supersteps, as demonstrated in \cite{MABDHLC10}.

\subsection{Connected Component}

We study two vertex-centric algorithms for the connected component problem --- hash-min and Shiloach-Vishkin (S-V),
considered in state-of-the-art literature \cite{YCXLNB14}.

\vspace{-1mm}
\subsubsection{Hash-Min Algorithm}

\footnotetext[2]{\scriptsize $K$ is \# iterations for convergence, $\alpha()$ functional inverse of Ackermann's function. $n_q$ and $m_q$ the number of nodes and edges, respectively, in the query graph.}

We assume that each vertex in a graph $G$ is assigned a unique ID.
The color of a connected component in $G$ is defined as the smallest
vertex among all vertices in the component. In Superstep 1,
each vertex $v$ initializes $min(v)$ as the smallest
vertex in the set $(\{v\} \cup neighbors(v))$, sends $min(v)$ to all $v$'s neighbors,
and votes to halt. In each subsequent superstep, a vertex $v$ obtains
the smallest vertex from the incoming messages, denoted by $u$. If
$u < v$, $v$ sets $min(v) = u$ and sends $min(v)$ to all its neighbors.
Finally, $v$ votes to halt. When all vertices vote to halt and there is no
new message in the network, the algorithm terminates.

\vspace{0.1mm}

It takes at most $\bigO(\delta)$ supersteps for the ID of the smallest vertex to
reach all the vertices in a connected component, and in each superstep, each vertex $v$
takes at most $\bigO(d(v))$ time to compute $min(v)$ and sends/receives
$\bigO(d(v))$ messages each using $\bigO(1)$ space. Therefore, it is a balanced
Pregel algorithm (i.e., satisfies properties 1-3), but not BPPA since the
number of supersteps can be larger than $\bigO(\log n)$, e.g., for a straight-line graph.

\vspace{0.1mm}

In each superstep, there are $\bigO(m)$ messages sent/received, and $\bigO(m)$ computations
are performed. By considering $g=\bigO(1)$; the time-processor product = $\bigO(m\delta)$.
This is more than the complexity of the best-known sequential algorithm,
which is due to BFS having complexity $\bigO(m+n)$.

\vspace{-1mm}
\subsubsection{Shiloach-Vishkin Algorithm}

\begin{figure}[tb!]
\centering
\subfigure [] {
\includegraphics[scale=0.15]{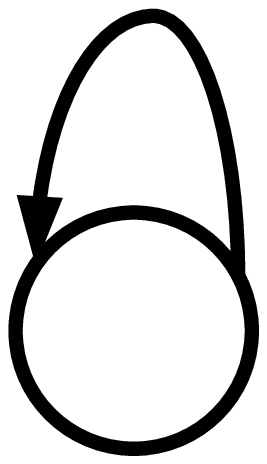}
\label{fig:SV1}
}
\subfigure [] {
\includegraphics[scale=0.15]{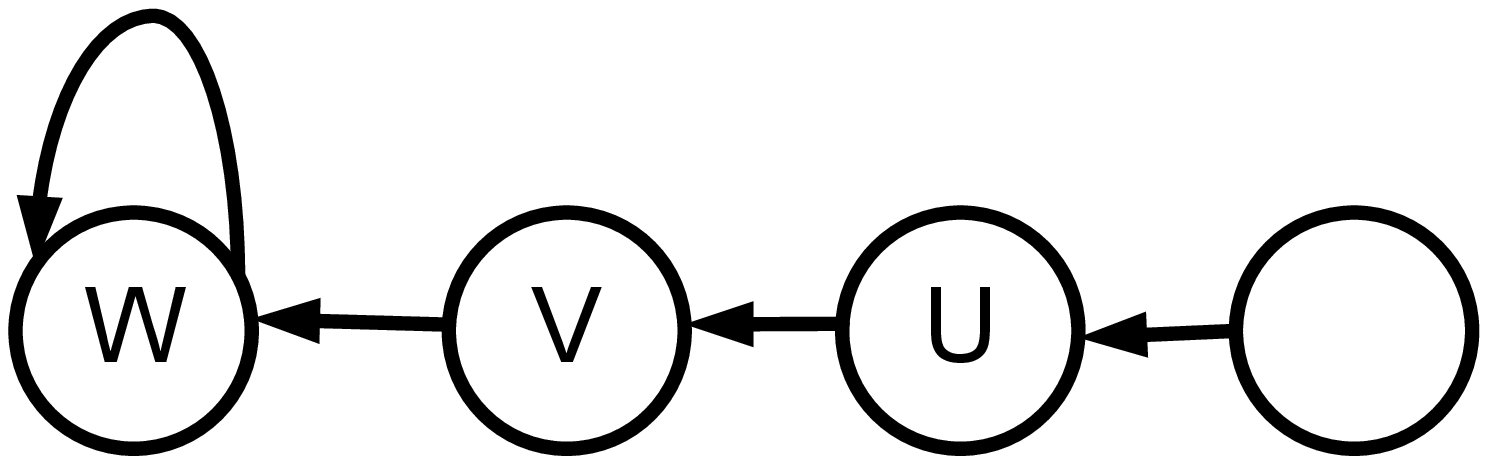}
\label{fig:SV2}
}
\subfigure [] {
\includegraphics[scale=0.15]{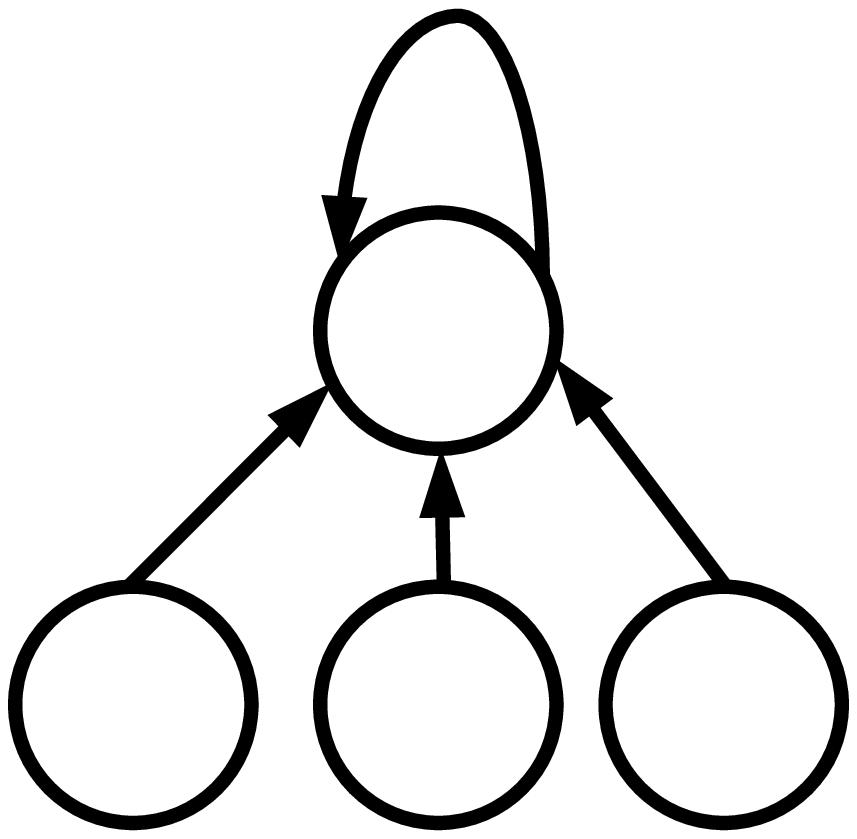}
\label{fig:SV3}
}
\vspace{-3mm}
\caption{\scriptsize Forest structure of S-V algorithm \cite{YCXLNB14}}
\label{fig:SV4}
\vspace{-5mm}
\end{figure}

In the S-V algorithm, each vertex $u$ maintains a pointer $D[u]$. Initially, $D[u] = u$,
forming a self-loop as depicted in Figure~\ref{fig:SV1}. During the algorithm, vertices are arranged
by a forest such that all vertices in each tree in the forest belong to the same connected component.
The tree definition is relaxed a bit to allow the tree root $w$ to have a self-loop
(see Figures~\ref{fig:SV2} and \ref{fig:SV3}), i.e., $D[w] = w$; while $D[v]$ of any other vertex $V$
in the tree points to $v$'s parent.

\vspace{0.1mm}

\begin{figure}[tb!]
\centering
\subfigure [\scriptsize Tree hooking] {
\includegraphics[scale=0.16]{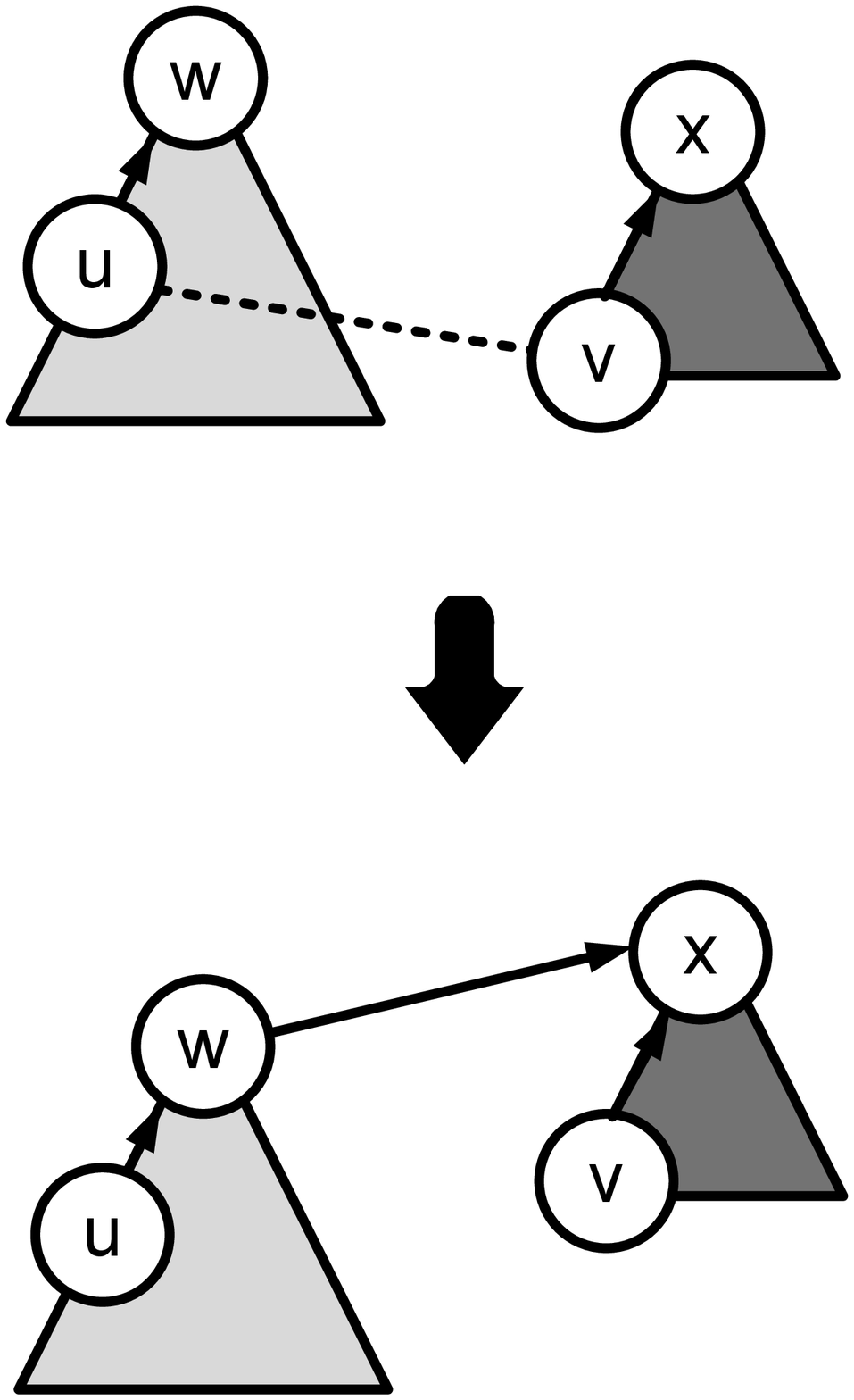}
\label{fig:SV5}
}
\subfigure [\scriptsize Star hooking] {
\includegraphics[scale=0.16]{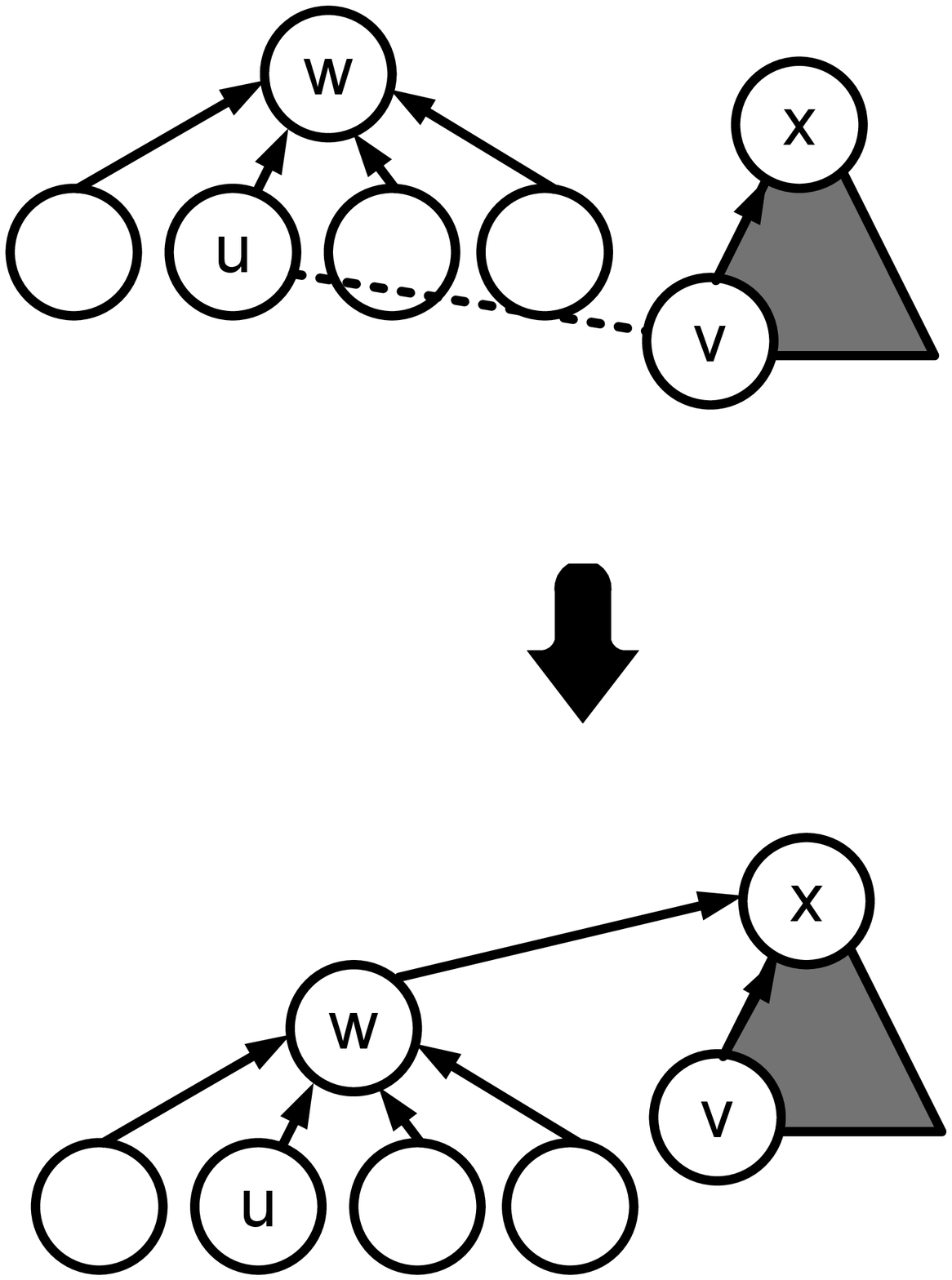}
\label{fig:SV6}
}
\subfigure [\scriptsize Shortcutting] {
\includegraphics[scale=0.16]{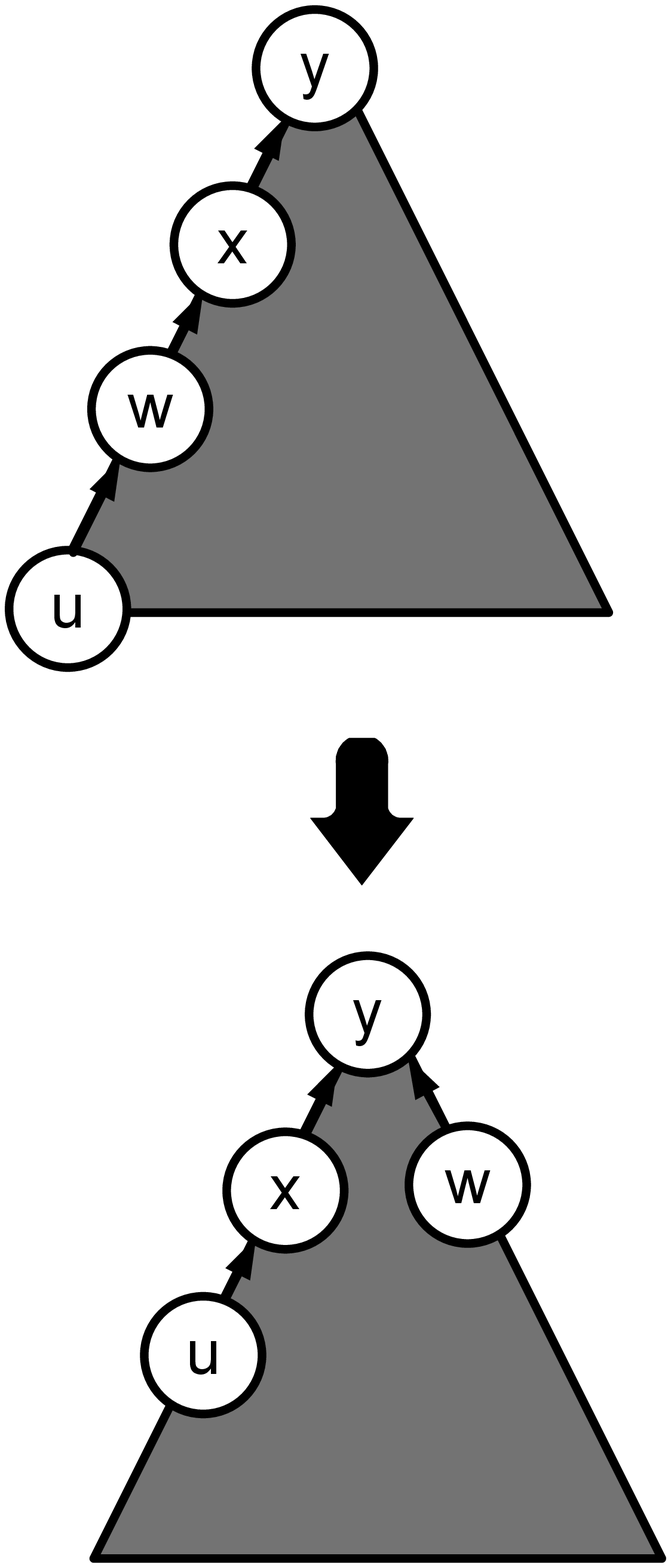}
\label{fig:SV7}
}
\vspace{-3mm}
\caption{\scriptsize Tree hooking, star hooking, and shortcutting \cite{YCXLNB14}}
\label{fig:SV8}
\vspace{-4mm}
\end{figure}

The S-V algorithm proceeds in iterations, and in each iteration, the pointers are updated in three steps
(Figure~\ref{fig:SV8}): (1) {\em tree hooking}: for each edge $(u, v)$, if $u$'s
parent $w = D[u]$ is a tree root, hook $w$ as a child of $v$'s parent $D[v]$
(i.e., merge the tree rooted at $w$ into $v$'s tree); (2) {\em star hooking}:
for each edge $(u, v)$, if $u$ is in a star (see Figure~\ref{fig:SV3} for an example of star),
hook the star to $v$'s tree as Step (1) does; (3) {\em shortcutting}: for each vertex $v$, move vertex $v$ and
its descendants closer to the tree root, by hooking $v$ to the parent of $v$'s parent, i.e., setting $D[v] = D[D[v]]$.
The algorithm terminates when every vertex is in a star. We perform tree hooking in Step (1) and
star hooking in Step (2) only if $D[v] < D[u]$, which ensures that the pointer values
monotonically decrease.

\vspace{0.1mm}

It was proved that the above S-V algorithm computes connected components in $\bigO(\log n)$ supersteps \cite{YCXLNB14}.
However, the algorithm is not a BPPA because a vertex $v$ may become the parent of more than $d(v)$
vertices and hence receives/sends more than $d(v)$ messages in a
superstep. On the other hand, the overall number of messages and computations in each superstep
are bounded by $\bigO(n)$ and $\bigO(m)$, respectively. With $g=\bigO(1)$, we
have the time-processor product = $\bigO((m+n)\log n)$. As earlier, this is higher than the complexity of the best-known
sequential algorithm.

\vspace{0.1mm}

For brevity, we omit the discussion on vertex-centric algorithms for weakly connected component, bi-connected component, and
strongly connected component. They can be found in \cite{YCXLNB14, SW14}. Since these methods use the vertex-centric connected component algorithm
(i.e., Hash-Min or S-V) as an underlying module, none of them are BPPA, and they perform more work than their best-known linear time
sequential algorithms (Table~\ref{tab:efficiency}).

\subsection{Tree Traversals}

\vspace{-2mm}
\subsubsection{Euler Tour}

\begin{figure}[tb!]
\centering
\subfigure [\scriptsize Euler tour \& pre-order] {
\includegraphics[scale=0.16]{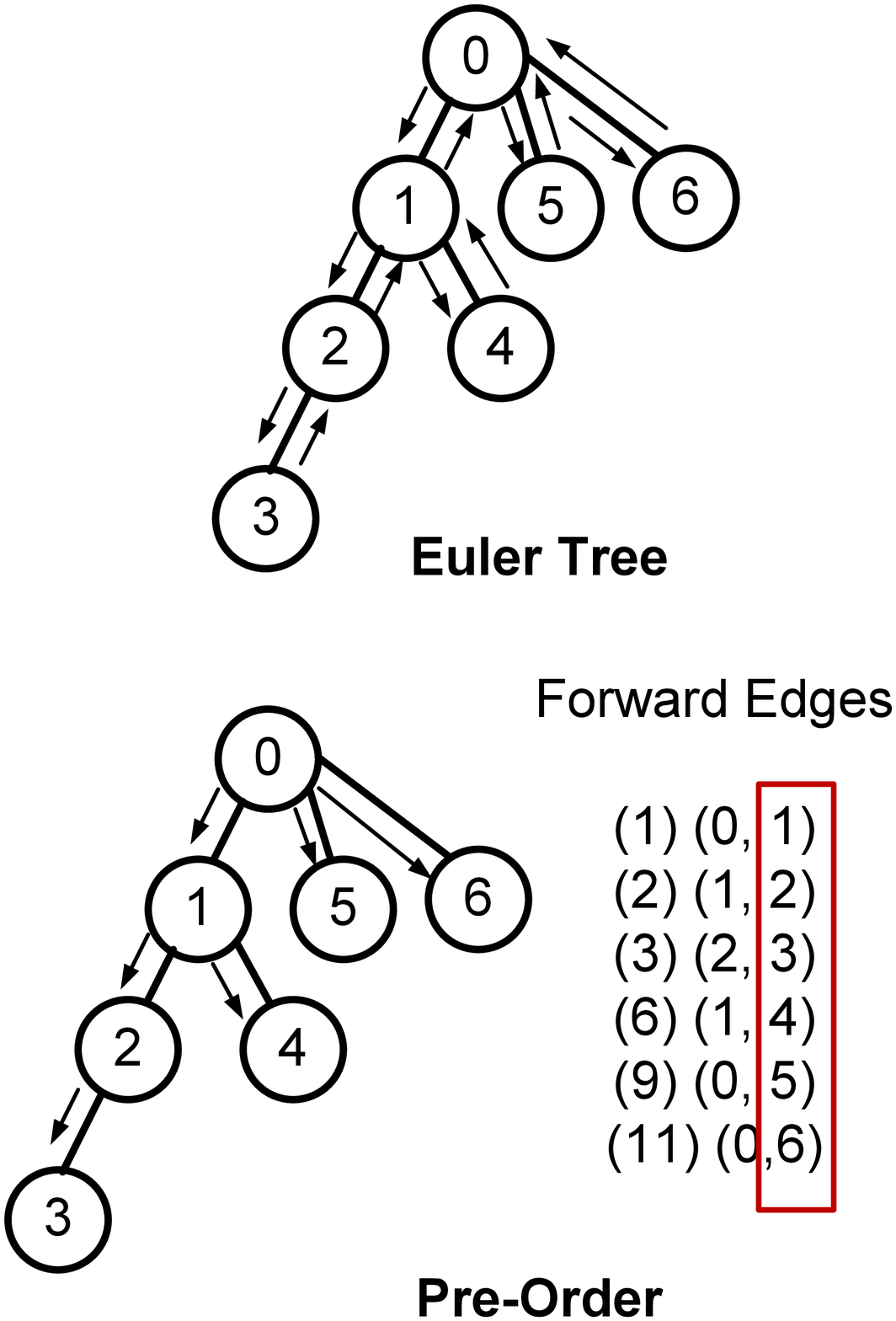}
\label{fig:euler}
}
\subfigure [\scriptsize List-ranking] {
\includegraphics[scale=0.16]{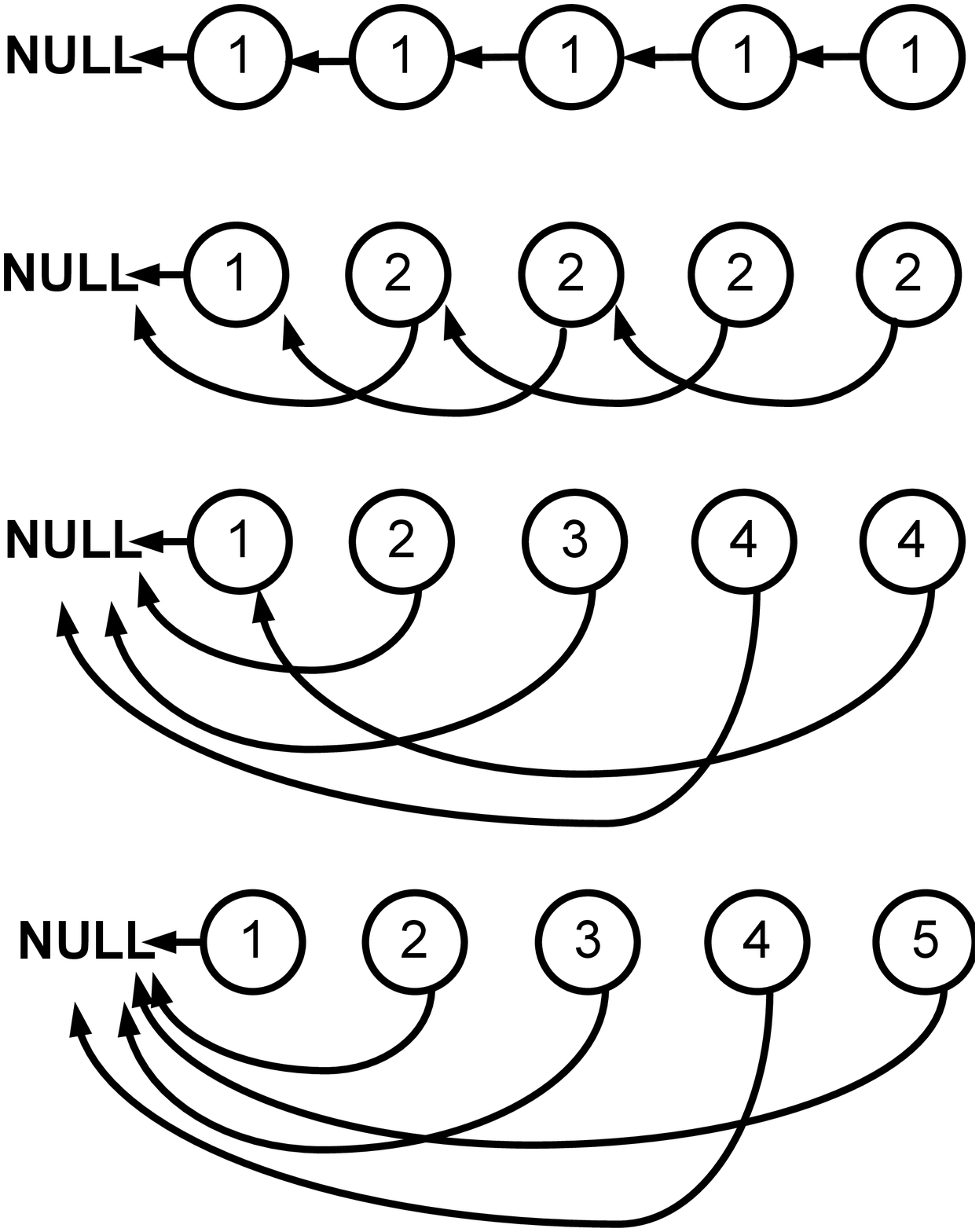}
\label{fig:list}
}
\vspace{-3mm}
\caption{\scriptsize Tree traversals and list-ranking}
\label{fig:tree}
\vspace{-4mm}
\end{figure}

A Euler tour is a representation of a tree, where each tree edge $(u, v)$ is considered
as two directed edges $(u, v)$ and $(v, u)$. As shown in Figure~\ref{fig:euler}, a Euler tour
of the tree is simply a Eulerian circuit of the directed graph, that is, a trail that visits every
edge exactly once, and ends at the same vertex where it starts.

\vspace{0.1mm}

We assume that the neighbors of each vertex v are sorted according to their IDs,
which is usually common for an adjacency list representation of a graph. For a vertex $v$, let
$first(v)$ and $last(v)$ be the first and last neighbor of $v$ in that sorted order; and for
each neighbor $u$ of $v$, if $u \ne last(v)$, let $next_v(u)$
be the neighbor of $v$ next to $u$ in the sorted adjacency list. We
also define $next_v(last(v)) = first(v)$. As an example, in Figure~\ref{fig:euler},
$first(0) = 1$, $last(0) = 6$, $next_0(1) = 5$, and $next_0(6) = 1$.

\vspace{0.1mm}

Yan et. al. \cite{YCXLNB14} designed a 2-superstep vertex-centric algorithm to construct the Euler tour as
given below. In Superstep 1, each vertex $v$ sends message $\langle u, next_v(u)\rangle$
to each neighbor $u$; in Supertep 2, each vertex $u$ receives the message $\langle u, next_v(u)\rangle$
sent from each neighbor $v$, and stores $next_v(u)$ with $v$ in $u$'s adjacency list.
Thus, for every vertex $u$ and each of its neighbor $v$, the next edge of $(u, v)$ is obtained as
$(v, next_v(u))$, which is the Euler tour.

\vspace{0.1mm}

The algorithm requires a constant number of supersteps. In
every superstep, each vertex $v$ sends/receives $\bigO(d(v))$ messages,
each using $\bigO(1)$ space. By implementing $next_v(.)$ as a hash
table associated with $v$, we can obtain $next_v(u)$ in $\bigO(1)$ expected
time given $u$. Therefore, the algorithm is BPPA. In addition, with $g=\bigO(1)$,
the time-processor product = $\bigO(n)$. This matches with the time complexity of the best-known sequential
algorithm.

\vspace{-2mm}
\subsubsection{Pre- and Post-Order Traversal}

The pre- and post-order numberings of the nodes
are obtained from Euler tour via a method called list-ranking as introduced below.
Let us consider a linked list $\mathcal{L}$ with $n$ elements,
where each element $v$ is associated with a value $val(v)$ and a
link to its predecessor $pred(v)$. However, the elements in $\mathcal{L}$
can be provided as input in any arbitrary order.
The element $v$ at the head of $\mathcal{L}$ has
$pred(v) = null$. For each element $v$ in $\mathcal{L}$, we define $sum(v)$ to be
the sum of the values of all the elements from $v$ following the predecessor
link to the head. The list-ranking problem computes $sum(v)$
for each element $v$.

\vspace{0.1mm}

A vertex-centric algorithm for list-ranking would be as follows (Figure~\ref{fig:list}). Initially, each vertex
$v$ assigns $sum(v) = val(v)$. Then, in subsequent rounds, each vertex $v$ does the following: If $pred(v) \ne null$, $v$
sets $sum(v) = sum(v) + sum(pred(v))$ and $pred(v) = pred(pred(v))$; otherwise, $v$ votes to halt.
This process repeats until $pred(v) = null$ for each vertex $v$; at this point, all vertices vote to halt and we have $sum(v)$
for all of them. The aforementioned list-ranking algorithm is BPPA because it terminates in $\bigO(\log n)$ supersteps, and each
element sends/receives at most one message per round. To compute the time-processor product, we note that
the element at position $i$ sends $\bigO(\log i)$ messages to its predecessors. Hence, the total
number of messages sent is $\bigO(\sum_{i=1}^n \log i)$, which is $\bigO(n\log n)$ due to  Stirling's approximation.
With $g=\bigO(1)$, the time-processor product = $\bigO(n\log n)$.

Next, let $pre(v)$ be the pre-order number of each vertex $v$ in the tree $T$. We compute pre-order numbers from the Euler tour
$P$ of the tree $T$ as follows (Figure~\ref{fig:euler}). We formulate a list-ranking problem by treating each edge $e \in P$ as a vertex and setting $val(e) = 1$.
After obtaining $sum(e)$ for each $e \in P$, we mark the edges in $P$ as forward/backward edges using a two-superstep BPPA.
In Superstep 1, each vertex $e = (u, v)$ sends $sum(e)$ to $e'=(v, u)$; in Superstep 2, each vertex $e'=(v, u)$ receives $sum(e)$
from $e = (u, v)$, sets $e'$ itself as a forward edge if $sum(e') < sum(e)$, and a backward edge otherwise.
To compute $pre(v)$, we run a second round of list-ranking by setting $val(e) = 1$ for each forward edge $e$ in $P$ and $val(e') = 0$
for every backward edge $e'$. Then, for each forward edge $e = (u, v)$, we get $pre(v) = sum(e)$ for vertex $v$. We set $pre(s) = 0$ for tree
root $s$. The post-order numberings can be obtained analogously by setting $val(e) = 0$ for each forward edge $e$ and $val(e') = 1$
for each backward edge $e'$ in $P$.

Finally, the proof that pre- and post-order computations are BPPA follows directly from the fact that both
Euler tour and list-ranking can be computed by BPPAs. However, due to list-ranking, the time-processor product
of this vertex-centric algorithm = $\bigO(n\log n)$, which is more than the complexity of the best-known sequential
algorithm (i.e., linear time with DFS) for the problem.

\subsection{Minimum Cost Spanning Tree}

Salihoglu et. al. implemented the parallel (vertex-centric) version of Boruvka's minimum cost
spanning tree (MCST) algorithm \cite{SW14,CC96} for a weighted, undirected graph $G$.
The algorithm iterates through the following phases, each time adding a set of edges to the MCST $S$ it constructs,
and removing some vertices from $G$ until there is just one vertex, in
which case the algorithm halts.

\vspace{0.1mm}

\begin{figure}[t!]
\centering
\includegraphics[scale=0.16]{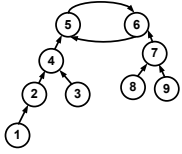}
\vspace{-3mm}
\caption{\scriptsize Conjoined-tree for MCST construction: vertex 5 is super-vertex}
\label{fig:mcst}
\vspace{-4mm}
\end{figure}

\noindent {\bf 1. Min-Edge-Picking:} In parallel, the edge list of each vertex is searched
to find the minimum weight edge from that vertex. Ties are broken by selecting the edge with minimum destination ID.
Each picked edge $(v, u)$ is added to $S$. As proved in \cite{CC96}, the vertices and their picked
edges form disjoint subgraphs $T_1,T_2, \ldots, T_k$, each of which
is a {\em conjoined-tree}, i.e., two trees, the roots of which are joined
by a cycle (Figure~\ref{fig:mcst}). We refer to the vertex with the smaller ID in the cycle of $T_i$ as
the super-vertex of $T_i$. All other vertices in $T_i$
are called its sub-vertices. The following steps merge all of the sub-vertices of every $T_i$
into the super-vertex of $T_i$.

\vspace{0.1mm}

\noindent {\bf 2. Super-vertex Finding:} First, we find all the super-vertices.
Each vertex $v$ sets its pointer to the neighbor $v$ picked in Min-Edge-Picking.
Then, it sends a message to $v$.pointer. If $v$ finds that it received a
message from the same vertex to which it sent a message earlier, it is part of the cycle. The
vertex with the smaller ID in the cycle is identified as the super-vertex.
After this, each vertex finds the super-vertex of the conjoined-tree it belongs to using the {\em Simple Pointer Jumping}
algorithm \cite{CC96}. The input $R$ to the algorithm is the set of super-vertices, and the input $S$
is the set of sub-vertices.

\vspace{1mm}

\noindent {\em Simple-Pointer-Jumping-Algorithm} $(R,S)$

\noindent $\quad$ {\bf repeat until} every vertex in $S$ points to a vertex in $R$

\noindent $\quad\quad$ {\bf for each} vertex $v$ that does not point to a vertex in $R$ {\bf do}

\noindent $\quad\quad\quad$ perform a pointer jump: $v$.pointer $\rightarrow$ $v$.pointer.pointer

\vspace{1mm}

\noindent {\bf 3. Edge-Cleaning-and-Relabeling:} We shrink each conjoined tree into the super-vertex
of the tree. This is performed as follows. In the set of edges of $G$, each vertex is renamed with the ID of the super-vertex of the conjoined
tree to which it belongs. The modified graph may have self-loops and multiple edges. All self-loops are removed. Multiple edges are removed
such that only the lightest edge remains between a pair of vertices.

One may verify that the above operations can be implemented in $\bigO(\delta)$ supersteps, which is due to the maximum number of iterations
required for the simple pointer jumping algorithm. Each superstep has message and computation complexity of $\bigO(m)$. The three above phases
are repeated, that is, the graph remaining after the $i$-th iteration is the input to the $i+1$-th iteration, unless it has just one vertex, in
which case the algorithm halts. Furthermore, the number of vertices of the graph at the $i+1$-th iteration is at most half of the number of vertices
at the $i$-th iteration. Hence, the number of iterations is at most $\bigO(\log n)$. With $g=\bigO(1)$, the time-processor
product = $\bigO(m\delta\log n)$. This is higher than the complexity of the best-known sequential algorithm for MCST, which is $\bigO(m\alpha(m,n))$ by
Chazelle's algorithm \cite{C00}. Here, $\alpha()$ is the classical functional inverse of Ackermann's function, and it
grows extremely slowly, so that for all practical purposes it may be considered a constant no greater than $4$. Even if we consider
a more widely-used Prim's algorithm (sequential), it has time complexity $\bigO(m+n\log n)$ using fibonacci heap and adjacency list.
In other words, the vertex-centric algorithm for MCST performs more work than the problem's sequential solutions.

\vspace{0.1mm}

The algorithm is not in BPPA, since (1) the Edge-Cleaning-and-Relabeling step increases the number
of neighbors of the super-vertices, and (2) the number of supersteps is $\bigO(\delta\log n)$.

\subsection{Graph Coloring}

The graph coloring problem deals with assigning colors to the vertices of a graph
such that adjacent vertices do not get the same color. The primary objective is to minimize the number
of colors used, which is \NP-hard. As there are several approximation and heuristic algorithms with different
performance guarantees, we study one of them --- graph coloring via maximal independent set (MIS), which was
implemented in the vertex-centric framework \cite{SW14}.  An MIS is a maximal set of vertices such that no pair
of vertices are adjacent. Luby's classic parallel algorithm \cite{L85} is used for iteratively finding an MIS
from the set of active vertices, assigns the vertices in the MIS a new color, and then removes them
from the graph, until no vertices are left in the graph.

\vspace{0.1mm}

Each iterative phase is processed as follows, where all vertices in the same MIS
are assigned the same color c: (1) each vertex $v$ is selected as a tentative vertex
in the MIS with a probability $\frac{1}{2\times d(v)}$; if a vertex has no neighbor
(i.e. an isolated vertex or becoming isolated after graph mutation), it is a trivial MIS; each tentative vertex
$v$ then sends $id(v)$ to all its neighbors; (2) each tentative vertex
$v$ receives messages from its tentative neighbors; let $min^*$ be
the smallest ID received, if $min^* > id(v)$, then $v$ is included in
the MIS and $color(v) = c$, and $id(v)$ is sent to its neighbors;
(3) if a vertex $u$ receives messages from its neighbors (that
have been included in the MIS in superstep (2)), then for each such
neighbor $v$, delete $v$ from $neighbors(u)$.

\vspace{0.1mm}

It was proved \cite{L85} that each iterative phase can be performed in expected $\bigO(\log n)$
supersteps, and each superstep has message and computation complexity $\bigO(m)$. Now, if there
are total $K$ iterative phases required for the graph coloring, the total number of supersteps
is $\bigO(K\log n)$. Usually, $K$ is not a constant, and in worst case, $K$ can be as large as
$\bigO(n)$ for a complete graph. Therefore, although it is a balanced Pregel algorithm,
this is not BPPA.

\vspace{0.1mm}

Following the above discussion, and with $g=\bigO(1)$,
the time-processor product = $\bigO(Km \log n)$. On the contrary, there exist $\bigO(m)$ time
maximal independent set finding algorithms, e.g., lexicographically first MIS. Therefore, sequential graph coloring via maximal independent set
can be computed in $\bigO(Km)$ time.

\subsection{Graph Simulation}

Graph simulation is a variant of the graph pattern matching problem,
which considers relations instead of functions from one node-labeled graph to another.
A graph $Q=(V',E',L')$ is said to be simulated by graph $G=(V,E,L)$ if there exists a binary relation $R$ between the nodes of
$Q$ and the nodes of $G$ such that:
(1) for each node $v$ in $Q$, there exists a node $u$ in $G$, such that $(v,u)\in R$, and
(2) for each node pair $(v,u) \in R$, (a) $L(v)= L(u)$, and (b) for each
edge $(v, v_1)$ in $Q$, there is an edge $(u,u_1)$ in $G$ such that $(v_1,u_1)$ is also in $R$, and $L(v,v_1)= L(u,u_1)$.

\vspace{0.1mm}

Fard et. al. implemented graph simulation over a vertex-centric framework \cite{FNRMS13}. A boolean flag, called $match$, is defined for
each vertex in $G$ in order to track if it matches a vertex in $Q$. At the
first superstep, the $match$ flag becomes true if its label matches the label of a vertex in $Q$.
In this case, a local match set, named $matchSet$, is generated to keep track of its potential
matches in $Q$. Next, each vertex  learns about the $matchSets$ of its children and keeps
them in a local list for later evaluation of graph simulation conditions (i.e., points 1 and 2 above).
Any match is removed from the local $matchSet$ if it does not satisfy the simulation conditions. The vertex
should also inform its parents about any changes in its $matchSet$. Consequently, any vertex that receives changes in
its children's $matchSet$s reflects those changes in its list of matched children and re-evaluates its own $matchSet$. The
algorithm can terminate after the third superstep if no vertex removes any match from its $matchSet$. Otherwise, this procedure
continues in superstep four and beyond until there is no change. It was shown that the the number of supersteps is upper
bounded by $\bigO(m)$. At the end, the local $matchSet$ of each vertex contains the correct and complete set of matches
between that vertex and the vertices of $Q$.

\vspace{0.1mm}

The algorithm is not BPPA because the number of supersteps can be asymptotically larger than $\bigO(\log n)$. Since the
$matchSet$ size for each vertex can be at most $\bigO(n_q)$, the message complexity at each superstep = $\bigO(mn_q)$.
The computation at each superstep = $\bigO(m(n_q+m_q))$. With $g=\bigO(1)$, we get the
time-processor product = $\bigO(m^2(n_q+m_q))$, which is more than the complexity of the best-known sequential algorithm
for graph simulation \cite{HHK95}.

\subsection{Difficult Graph Problems for \\ Vertex-Centric Model}

Since computations in vertex-centric model happen at vertex level, an important question would be whether {\em all} kinds
of graph analytics tasks and algorithms can be expressed {\em efficiently} in this framework.
(1) Vertex-centric model usually operates on the entire graph, which is often not necessary for online ad-hoc queries \cite{KE14},
including shortest path, reachability, and subgraph isomorphism. (2) This model is not well-suited for graph analytics
that require a subgraph-centric view around vertices, e.g., local clustering coefficient, triangle and motifs counting. This
is due to the communication overhead, network traffic, and the large amount of memory required to construct multi-hop neighborhood
in each vertex's local state \cite{QDL14}. (3) Not all distributed algorithms for the same graph problem can be implemented
in a vertex-centric framework. As an example, it is difficult to implement the distributed union-find algorithm for the connected
component problem using a vertex-centric model \cite{MIM15}. However, this algorithm is useful for edge-streams. (4) State-of-the-art research on vertex-centric graph processing
mainly focused on a limited number of graph workloads such as PageRank and connected components, and it is largely unknown whether
some other widely-used graph computations, e.g., modularity optimization for community detection,
betweenness centrality (weighted graphs), influence maximization, link prediction, partitioning, and
embedding can be implemented efficiently over vertex-centric systems.

\section{Discussion and Conclusion}
\label{sec:conclusion}

It is difficult to express many graph problems and algorithms in
a vertex-centric model. Even for the ones that were implemented in state-of-the-art literature, our benchmark
shows that they often suffer from imbalanced workload/ large number of iterations, and perform more work than their best
known sequential algorithms. Due to such difficulties, alternate proposals exist where the entire
graph is loaded on a single machine having larger memory, or on a multi-core machine with shared-memory.
Nevertheless, distributed graph processing systems would still be critical due to the two following reasons.
First, graph analysis is usually an intermediate step of some larger data analytics pipeline, whose previous and following
steps might require distribution over several machines. In such scenarios, distributed graph processing
would help to avoid expensive data transfers. Second, distributed-memory systems generally scale well,
compared to their shared-memory counterparts.

\vspace{0.1mm}

However, one distributed model might not be suitable for all kinds of graph computations. Many recent distributed systems, e.g.,
Trinity, NScale, and Apache Flink support multiple paradigms,
including vertex-centric, subgraph-centric, stream dataflow, and shared access. But, perhaps more
importantly, we need to identify the appropriate metrics to evaluate these systems. In addition to
time-processor product and BPPA that we studied in this work, one can also investigate
the speedup and cost/computation. Two other critical metrics are {\em expressibility} and
{\em usability}, which were mostly ignored due to their qualitative nature. The former identifies
the workloads that can be efficiently implemented in a distributed framework, while the later
deals with ease in programming, e.g., domain-specific languages, declarative
programming, high-level abstraction to hide data partitioning, communication, system architecture, and
fault tolerance, as well as availability of debugging and provenance tools. With all these exciting
open problems, this research area is likely to get more attention in the near future.   

\vspace{-2mm}
{\scriptsize
\bibliographystyle{abbrv}
\bibliography{ref}
}


\end{document}